\documentclass[11pt,a4paper]{article}
\pdfoutput=1
\usepackage{a4wide}
\usepackage{amsfonts}
\usepackage{amssymb}
\usepackage{amsmath}
\usepackage{graphicx}
\usepackage{booktabs}
\usepackage{array}
\usepackage{color}
\usepackage{ulem}
\usepackage[small,bf]{caption}
\usepackage{cite}
\usepackage[usenames,dvipsnames]{xcolor}
\usepackage{subfigure}
\usepackage{verbatim}

\allowdisplaybreaks

\numberwithin{equation}{section}

\newcounter{mysubequation}[equation]

\DeclareMathOperator{\ci}{\text{i}}
\DeclareMathOperator{\diff}{\text{d}}

\begin{document}
\begin{titlepage}

\vspace*{-15mm}
\begin{flushright}
NCTS-PH/2009
\end{flushright}
\vspace*{0.7cm}

\begin{center}
{
\bf\LARGE Light Dark Matter Scattering in\\[0.3em]
Gravitational Wave Detectors
}
\\[8mm]
Chun-Hao~Lee$^{\, a,}$ \footnote{E-mail: \texttt{lee.chunhao9112@gmail.com}},
Chrisna~Setyo~Nugroho$^{\, b,}$ \footnote{E-mail: \texttt{csnugroho@cts.nthu.edu.tw}} and 
Martin~Spinrath$^{\, a,}$ \footnote{E-mail: \texttt{spinrath@phys.nthu.edu.tw}}
\\[1mm]
\end{center}
\vspace*{0.50cm}
\centerline{$^{a}$ \it Department of Physics, National Tsing Hua University, Hsinchu, 30013, Taiwan}
\vspace*{0.2cm}
\centerline{$^{b}$ \it Physics Division, National Center for Theoretical Sciences,}
\centerline{\it National Tsing-Hua University, Hsinchu, 30013, Taiwan}
\vspace*{1.20cm}

\begin{abstract}
\noindent
We present prospects for discovering dark matter scattering in
gravitational wave detectors.
The focus of this work is on light, particle dark matter with 
masses below 1 GeV/$\text{c}^{2}$.
We investigate how a potential signal compares
to typical backgrounds like thermal and quantum noise,
first in a simple toy model and then using KAGRA as 
a realistic example.
That shows that for a discovery much lighter and cooler mirrors
would be needed.
We also give some brief comments on space-based experiments and
future atomic interferometers.
\end{abstract}

\end{titlepage}
\setcounter{footnote}{0}

\section{Introduction}

The nature of Dark Matter (DM) is one of the very
important issues in fundamental physics. It affects
particle physics, astrophysics and cosmology all at once.
In fact, particle physicists for a long time suggested
candidates with a mass in the GeV/c$^2$ to TeV/c$^2$
range motivated by the hierarchy problem. This can be solved
elegantly by adding new particles in this mass range,
some of which can play the role of DM.

So far direct searches in dedicated DM experiments
and the LHC did not provide any conclusive evidence
for any DM candidates around the weak scale.
That is one of the reasons, why recently there have been 
stronger efforts to increase the sensitivity towards lighter 
and heavier DM masses, see, e.g., the community 
report~\cite{Alexander:2016aln}.

On the other hand, with the discovery of gravitational waves (GWs) by LIGO and
VIRGO in 2016~\cite{Abbott:2016blz} particle physicists also
developed a growing interest in this technology. One of the authors
discussed with other collaborators the prospects 
for the discovery of the Cosmic Neutrino Background using laser
interferometers
also briefly mentioning DM in~\cite{Domcke:2017aqj}. Scattering of
DM particles in GW detectors and interferometers in general
can lead to an effect similar to Brownian motion
which was discussed from a particle physicist's point
of view in~\cite{Cheng:2019vwy}. 
Shortly after, a similar idea was discussed in \cite{Tsuchida:2019hhc},
which employs a more common language in the GW community.
The main motivation of this paper is to
build a framework which realistically models
a DM scattering signal in an optomechanical setup like a GW detector.
Within this framework we can then
compare a potential
DM signal to common backgrounds in GW detectors in the language
of the GW community. 
As an explicit and realistic example we will consider KAGRA~\cite{Somiya:2011np}
and also briefly comment on other technologies.

Actually, the aforementioned works have not been the only
proposals to search for DM at GW detectors
or interferometers in general, 
see, for instance,~\cite{Seto:2004zu, Adams:2004pk, Riedel:2012ur, PhysRevLett.114.161301,
	Arvanitaki:2015iga, Stadnik:2015xbn, Branca:2016rez, Riedel:2016acj, Hall:2016usm,
	Jung:2017flg, Pierce:2018xmy, Morisaki:2018htj, Grote:2019uvn, Badurina:2019hst, 
	Graham:2017pmn, Coleman:2018ozp, Bertoldi:2019tck, Dimopoulos:2008sv, 
	Graham:2012sy, Graham:2016plp, Canuel:2017rrp, Canuel:2019abg, Zhan:2019quq, Dimopoulos:2008hx, Kawasaki:2018xak, Monteiro:2020wcb}.
Nevertheless, our approach here differs from these works.
They usually focus on very light DM which has wave-like properties
or very heavy DM which produces a GW signal itself.
We instead focus here on light DM in the mass range between
1~keV/c$^2$ and 1~GeV/c$^2$ where DM has a particle nature and our
DM signal is modelled as an elastic scattering event with components
of the interferometer.

Our paper is organised as follows: In section~\ref{sec:toymodel}
we first discuss a damped harmonic oscillator as a  
simplified toy model for the mirror system in realistic GW detectors. For that model we discuss the most important sources
of noise and a potential DM signal, from which we can also derive
simple formulas for the signal-to-noise ratio (SNR) which allows to
provide a rough estimate for
the DM sensitivity of any terrestrial GW detector.
In section~\ref{sec:KAGRA} we generalise the formulas derived
in section~\ref{sec:toymodel} and apply them to the concrete
case of the KAGRA detector. In section~\ref{sec:Others} we 
briefly comment on space-based experiments and atomic interferometers
before we summarise and conclude in section~\ref{sec:Summary}.

\section{Simplified Setup: Damped Harmonic Oscillator}
\label{sec:toymodel}

Before we discuss KAGRA as a realistic example we would like to first
study how a potential DM signal
compares to thermal and quantum noise for a toy model of a damped harmonic
oscillator in one dimension. The calculations here can later be easily extended to the KAGRA case.

We base our discussion on 
the complex differential equation
\begin{align}
\label{eq:cDEQ}
  m \, \ddot{x}_c  + k_{c} \, (1+ \ci \phi) \, x_c = \frac{F_{\text{ext},c}}{L}  \;,
\end{align}
where $x_c(t)$ is the dimensionless relative displacement in complex form,
$L$ is the interferometer arm length, $m$ is the mass of the oscillator (the test mirror (TM))
and $k_c$ is the equivalent to a spring constant of the suspension system.
The suspension loss angle, $0 < \phi \ll 1$, describes the damping
of the system.
$F_{\text{ext},c}$ is an external force applied to the system which will be
adjusted depending on which physics aspect we are discussing.
This equation has the form of how KAGRA published the parameters
describing their suspension system.

This differential equation has only complex solutions due to its complex coefficients,
which has advantages and disadvantages. One of the big advantages is that it
describes the damping induced by internal friction well, cf.~\cite{Nowick:1972}. This internal friction is
frequency dependent
which makes it difficult to write down explicitly for a system of coupled
oscillators as in KAGRA.
A disadvantage is, that
the displacement $x_c$ does not correspond immediately to an observable quantity which
makes the modelling of the external force less intuitive.

If we would want to describe the same system instead in terms of a 
more intuitive real differential equation describing a real 
displacement we can use
\begin{equation}
 \label{eq:rDEQ}
  m \, \ddot{x}_{r} + 2 \, m \, \omega_r \, \xi \, \dot{x}_{r} + m\, \omega_r^2 \left(1 + \xi^2 \right) x_{r} = \frac{F_{\text{ext},r}}{L}  \;,
\end{equation}
and we will discuss later how we match the quantities in this equation with
the quantities in eq.~\eqref{eq:cDEQ}.

In our analysis we follow a standard approach for GW detectors described
in \cite{Moore:2014lga}. In this reference, the measured output of the detector
is split into a noise and a signal part. For the toy model considered here, the total displacement 
\begin{equation}
x_{\text{tot},c}(t) = x_{\text{th},c}(t) + x_{\text{qu},c}(t) + x_{\text{DM},c}(t)
\end{equation}
is given by the superposition of thermal noise $x_{\text{th},c}(t)$, quantum noise $x_{\text{qu},c}(t)$
and the signal displacement induced by DM scattering $x_{\text{DM},c}(t)$ and similarly for the
real displacement.
In a realistic experiment there are other sources of noise but here we want to focus on the
most sensitive region of GW detectors, KAGRA in particular, where the suspension thermal noise
and quantum noise are mostly dominant, cf.~\cite{KAGRAnoise}.

\subsection{Noise}
\label{subsec:toymodelthermalnoise}

First, we discuss the suspension thermal noise, following closely the approach described
in \cite{Saulson:1990jc,Saulson:2017} based on the
fluctuation-dissipation theorem \cite{Callen:1951,Callen:1952}. Accordingly, 
the one-sided power spectral density (PSD) of the thermal noise $S_{\text{th}}$,
is given by
\begin{equation}
\label{eq:psd_noise}
S_{\text{th}}(\omega)=\frac{4 \, k_B \, T}{L^2 \, \omega^2} \Re[Y(\omega)] \;,
\end{equation}
where $k_B$ is Boltzmann's constant, $T$ the temperature of the system,
$L$ the interferometer arm length
as before, $\omega$ the frequency of the external force, and $\Re[Y(\omega)]$
is the real part of the complex admittance
in the frequency domain.

The admittance is defined for periodic external forces. For a given frequency,
$\omega$, of the external force, i.e.~$F_{\text{ext}} \sim \text{exp}(\ci \omega \, t)$,
the time dependence of the steady state solution would be of the form
$x_{\text{th},c}(t) \sim \text{exp}(\ci \omega \, t)$
as well. The admittance for the complex case is then given by
\begin{equation}
 Y_c(\omega) \equiv \frac{\dot{x}_{\text{th},c}}{F_{\text{ext}}/( m \, L)} = \text{i} \, \omega \, D_c^{-1}(\omega) =\frac{\ci \omega  }{  (- m \, \omega^2 + k_c (1 + \ci \phi) )   } \;,
\label{eq:YcEQ}
\end{equation}
where we have introduced $D_c(\omega)$ as the Fourier transform of the
differential operator, in this case
$D_c(\omega) = - m \,\omega^2 + k_c (1 + \ci \phi)$.
The real part of the associated admittance is
\begin{equation}
 \Re[Y_c(\omega)]  = \Re \left[ \frac{\ci \omega  (- m \, \omega^2 + k_c - \ci k_c \, \phi ) }{  (k_c - m \, \omega^2)^2 +  k_c^2  \phi^2   } \right] = \frac{k_c \, \omega \, \phi}{(k_c - m \, \omega^2)^2 +  k_c^2  \phi^2 } \;.
\end{equation}
The corresponding thermal noise PSD originated from the internal damping of the system is therefore given by
\begin{align}
\label{eq:cThPSD}
S_{\text{th},c}(\omega) &= \frac{4 \, k_{B} \, T}{L^2}  \frac{k_c \, \phi/\omega}{ (k_c - m \, \omega^2)^{2} + \, k_c^2 \, \phi^2 } = \frac{4 \, k_B \, T \,}{L^2} \frac{ \phi \, \omega_c^2/(m \,\omega)}{  (\omega^2 - \omega_c^2)^2 + \omega_c^4 \phi^2  } \;,
\end{align}
where we have introduced $\omega_c^2 \equiv k_c/m$. Notice that the PSD vanishes
when the loss angle $\phi$ equals to zero.
Following the same procedure, one obtains the thermal noise PSD from the real
differential eq.~\eqref{eq:rDEQ},
cf.~\cite{Saulson:1990jc}, as
\begin{align}
\label{eq:rThPSD}
 S_{\text{th},r}(\omega) &= \frac{4 \, k_{B} \, T}{L^2}   \frac{2 \, \omega_r \, \xi/m }{ (\omega^2 - \omega_r^2 \left(1 + \xi^2 \right) )^2 + (2 \, \omega_r \, \xi)^2 \omega^2 } \nonumber \\
  &= \frac{4 \, k_{B} \, T}{L^2}   \frac{2 \, \omega_r \, \xi/m }{ (\omega^2 - \omega_r^2 \left(1 - \xi^2 \right) )^2 + 4 \, \omega_r^4 \, \xi^2 } \;.
\end{align}

Now one can argue that since both equations describe the same
physical system the thermal noise spectra $S_{\text{th},c}$ and 
$S_{\text{th},r}$ should be identical, i.e.~$S_{\text{th},c} = 
S_{\text{th},r}$.
From this we derive relations between the parameters in eq.~\eqref{eq:cDEQ}
and eq.~\eqref{eq:rDEQ} as 
\begin{equation}
\begin{split}
\label{eq:thMatching}
 \omega_r^2 &= \omega_c^2 \left( 1 - \frac{\omega_c^2}{\omega^2} \frac{\phi^2}{4} \right)   \;, \\
 \xi &= \frac{\omega_c}{ \omega } \left( 1 - \frac{\omega_c^2}{\omega^2} \frac{\phi^2}{4} \right)^{-1/2} \frac{\phi}{2} \approx \frac{\omega_c}{ \omega }  \frac{\phi}{2} \left( 1 + \frac{\omega_c^2}{\omega^2} \frac{\phi^2}{8} \right) \;.
 \end{split}
\end{equation}
Here we see the forementioned dependence of $\xi$ on the frequency
of the driving force explicitly which has been observed in systems
with internal damping. On a side-note, in KAGRA and LIGO ordinary 
viscous damping with a constant $\xi$ induced by surrounding gas is 
negligible, cf.~\cite{Saulson:1990jc}.

In the next subsection, we present  an alternative way to determine
$\omega_r$ and $\xi$ which is more immediately related to the physical
situation where we need them. For that reason we will not use the 
matching from eq.~\eqref{eq:thMatching} explicitly later on. 
Nevertheless, both matchings differ only in $\mathcal{O}(\phi^2)$ 
which is negligibly small for KAGRA.
Indeed, it is long known that it is not possible
to match a complex and a real differential equation for a system
with internal damping in a uniquely consistent way~\cite{Neumark:1957}.

In the following discussion, we always refer to the thermal noise
derived from the complex equation: $S_{\text{th}} \equiv S_{\text{th},c}$.
The readout of the GW experiment is expressed in terms of the
strain amplitude which is related to the PSD as \cite{Moore:2014lga}
\begin{align}
\label{eq:strainthermal}
h_{\text{th}}(\omega) = \sqrt{ S_{\text{th}}(\omega) } \;.
\end{align}

Next, for the quantum noise induced by the uncertainty principle of the measurement
we use in the toy model the so-called standard quantum limit (SQL) \cite{SQL}
\begin{equation}
 S_{\text{qu}} = \frac{8 \, \hbar}{ m \, \omega^2 \, L^2 } \;.
  \label{eq:sSQL}
\end{equation}
Modern GW detectors can do better than this in certain frequency ranges,
see, e.g.,~\cite{Buonanno:2001cj,Kimble:2000gu}. The quantum noise in such
setups depends on a lot of parameters though and we want to keep
things simple and transparent here.

The strain of the quantum noise is
again related to the PSD via the standard relation \cite{Moore:2014lga}
\begin{equation}
 h_{\text{qu}}(\omega) =  \sqrt{ S_{\text{qu}}(\omega) } \;.
 \label{eq:SQL}
\end{equation}

Finally, the total noise PSD of the toy model is given by the thermal
noise and quantum noise
\begin{equation}
 S_n =   S_{\text{th}} + S_{\text{qu}}  \;,
 \label{eq:SnToy}
\end{equation}
with the corresponding strain amplitude
\begin{equation}
h_n = \sqrt{  h_{\text{th}}^2 + h_{\text{qu}}^2 } \;.
\label{eq:hToy}
\end{equation}

\subsection{DM signal}
\label{sec:toyDM}

In this subsection, we focus on the signal part of the output induced by the DM hit.
We consider the situation where the TM is being hit by a DM particle
at $t=0$ transferring a recoil momentum $q_R$ to it.
This picture can be straightforwardly implemented into the real
differential eq.~\eqref{eq:rDEQ} setting the external DM force to
$F_{\text{ext},r} = q_R \, \delta(t)$,  cf.~\cite{Tsuchida:2019hhc}.

The mechanical loss of the KAGRA components were actually tested using a
pulse-like force~\cite{Chen:2016} exactly what we assume for the DM signal.
The response of the system was fitted to a function
\begin{equation}
\label{eq:rsolu}
 x_{r}(t) \sim \text{exp}\left(- \omega_r \, \xi \, t \right) \sin (\omega_r \, t) \;,
\end{equation}
where we have slightly adapted the notation of Chapter 5 of \cite{Chen:2016}
to our notation such that it solves the free version of eq.~\eqref{eq:rDEQ}
with no external force.

We now compare this solution to the free solution 
of eq.~\eqref{eq:cDEQ} which has the form 
\begin{equation}
 x_{\text{DM},c}(t) \sim \exp(- \omega_c \, \Im(\sqrt{1 + \ci \phi}) \, t ) \exp(\ci \omega_c \, \Re(\sqrt{1 + \ci \phi}) \, t) \;.
\end{equation}
Since both eq.~\eqref{eq:cDEQ} and eq.~\eqref{eq:rDEQ} should describe the
same (free) physical system we demand that the damping of the amplitude
and the oscillation frequency in both cases are identical. From that we
determine the relations between the parameters in both cases as
\begin{equation}
\begin{split}
\label{eq:fMatching}
 \omega_r &= \omega_c \, \Re(\sqrt{1 + \ci \phi}) = \omega_c \left( 1 + \phi^2 \right)^{1/4} \cos\left( \frac{1}{2} \arctan \phi\right) \approx \omega_c \left( 1 + \frac{\phi^2}{8} \right)  \;,\\
 \xi &= \frac{\Im(\sqrt{1 + \ci \phi})}{\Re(\sqrt{1 + \ci \phi})} = \tan\left( \frac{1}{2} \arctan \phi\right) \approx \frac{\phi}{2} \left( 1 - \frac{\phi^2}{4} \right) \;.
 \end{split}
\end{equation}
We see that up to $\mathcal{O}(\phi^2)$ this solution is identical to
the leading order one from
the matching of the thermal spectra in eqs.~\eqref{eq:thMatching} 
after setting $\omega = \omega_r$. In fact, this is the only 
available choice
for $\omega$ since we consider here a free oscillator where the only
available physical frequency is the oscillation frequency $\omega_r$.

We further assume the initial condition $x_{\text{DM},r}(t) = 0$
for $t \leq 0$ and we model only one hit at $t=0$. For a given recoil
momentum $q_R$, the dimensionless displacement due to the DM hit is 
then given by
\begin{align}
 x_{\text{DM}}(t) &= \theta(t) \frac{q_R}{ m \, \omega_r \, L} \text{exp}\left(- \omega_r \, \xi \, t \right) \sin (\omega_r \, t) \;,
\end{align}
where we have dropped the index $r$ from the displacement.

In order to calculate the strain amplitude,
we need the Fourier transform of the displacement and its respective modulus squared
\begin{align}
\label{eq:rxtDM}
 \tilde{x}_{\text{DM}}(\omega) &= \int \diff t \, x_{\text{DM}}(t) \, \text{e}^{-\ci \omega \, t} = \frac{q_R}{ m  \, L} \frac{1}{\omega_r^2 - (\omega - \ci \omega_r \, \xi)^2} \;, \\
 \label{eq:rxtDMSq}
 | \tilde{x}_{\text{DM}}(\omega) |^2 &= \frac{q_R^2}{ m^2  \, L^2} \frac{1}{  \left( \omega^2 - \omega_r^2 (1 - \xi^2) \right)^2 + 4 \, \omega_r^4 \, \xi^2 } \;.
\end{align}
It is easy to see here that if we would have multiple hits in the observed 
time window the result would be simply
\begin{align}
\tilde{x}_{\text{DM}}(\omega) &= \frac{1}{ m  \, L} \frac{1}{\omega_r^2 - (\omega - \ci \omega_r \, \xi)^2} \sum_i q_R^{(i)} \text{e}^{-\ci \omega \, t_i} \;,
\end{align}
where $q_R^{(i)}$ is the recoil momentum (with either sign) of the $i$-th
hit at time $t_i$. In the analysis, it would then be recommended to choose
different time intervals for the Fourier transform to determine the times
when a hit occurred. In fact, random hits as we study them here are more
easily studied and analysed in the time-domain, cf.~\cite{Cheng:2019vwy}.
The advantage of using the frequency-domain approach is the comparability
to the results from GW detectors where everything is presented in this way.
To keep things
simple, we will in the following only consider the case of one hit at $t=0$.

The DM induced strain amplitude is given by \cite{Moore:2014lga}
\begin{equation}
\label{eq:toyDMstrain}
 h_{\text{DM}}(\omega) =  \sqrt{\frac{2 \, \omega}{\pi}} \,  |\tilde{x}_{\text{DM}}(\omega)| \;.
\end{equation}

\begin{figure}
	\centering
	\includegraphics[width=0.8\textwidth]{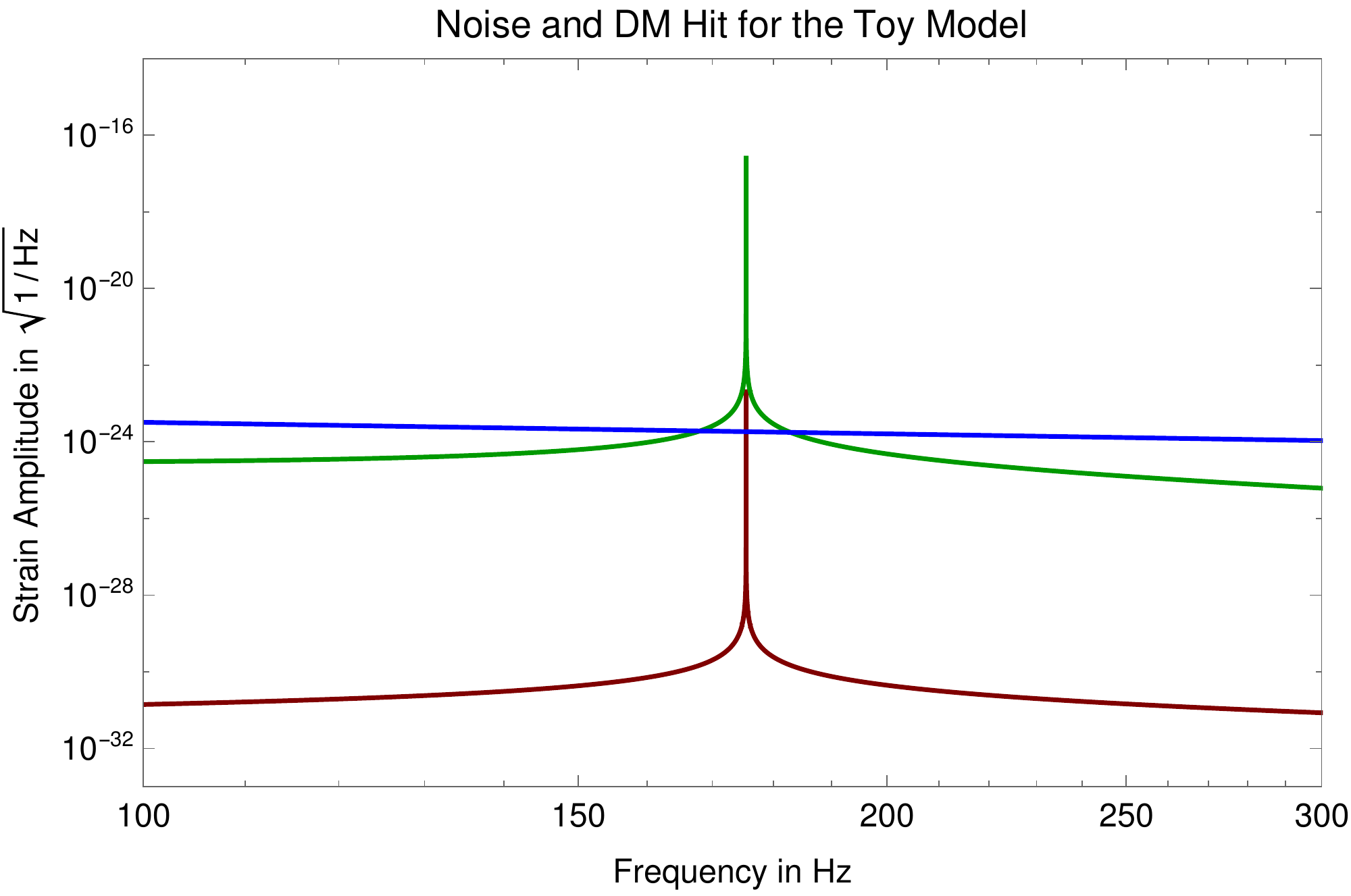}
	\caption{The strain amplitudes in our toy model for the
		thermal suspension noise 
		(green line), the quantum noise (blue line) and a hypothetical DM 
		signal (maroon line). For more details, see main text. 
		}	
	\label{fig:ToyModel}
\end{figure}

In Fig.~\ref{fig:ToyModel}, we plot the strain amplitudes for the DM 
signal and the noise for the toy model.
The noise input parameters are specified as:
$f_0 = \omega_c/(2 \, \pi) = 175.4$~Hz, $\phi = 6.32 \times 10^{-12}$, 
$m = 22.8$~kg, $T = 19$~K, $L = 3$~km.
The frequency $f_0$ corresponds to one of the peaks in the thermal suspension
noise of KAGRA and the loss angle $\phi$ was chosen to correspond to the
width of that peak. The other parameters are chosen to agree with KAGRA
as well.
The DM recoil momentum is roughly approximated by the DM momentum
$q_R = m_{\text{DM}} \, |\vec{v}_{\text{DM}}|$, where we use
$m_{\text{DM}}= 1$~GeV/$\text{c}^2$ and 
$|\vec{v}_{\text{DM}}|=220$~km/s.
From Fig.~\ref{fig:ToyModel} we see that a potential light DM signal at 
KAGRA is pretty much hidden behind the noise, the thermal
suspension noise in particular. The peak of the
thermal noise nevertheless coincides with the peak of the DM signal,
while off-resonance they behave differently.
For smaller frequencies the DM signal falls faster while for large 
frequencies the thermal noise falls faster.
To give a more quantitative statement
regarding the DM signal, we will discuss the SNR
in the next subsection.

But before we discuss this, we present an alternative approach to derive 
$|\tilde{x}_{\text{DM}}(\omega)|^2$ directly from the complex
differential eq.~\eqref{eq:cDEQ} without having to go
through the matching to a real differential equation as in
eq.~\eqref{eq:rDEQ} making the realistic considerations for KAGRA
much easier.

We start with the Fourier expansion of the dimensionless displacement 
induced by DM as
\begin{align}
 x_{\text{DM}} (t) \equiv \int \diff \omega \, \tilde{x}_{\text{DM}} (\omega) \exp(\ci \omega \, t) \;,
\end{align}
and we set the external force for the complex case equal to the external
force for the real case
\begin{align}
F_{\text{ext},c} = F_{\text{ext},r} = q_R \, \delta(t) = q_{R} \, \int \diff \omega \,  \exp(\ci \omega \, t) \;.
\end{align}
The Fourier transformed displacement $\tilde{x}_{\text{DM}}(\omega)$ can then
be obtained easily after taking the Fourier transform of eq.~\eqref{eq:cDEQ}
\begin{equation}
 \left( - \omega^2 + \omega_c^2 (1 + \ci \phi) \right)\tilde{x}_{\text{DM}}(\omega) = \frac{q_R}{m \, L} \;.
 \label{eq:FouriercDEQ}
\end{equation}
The corresponding absolute square of $\tilde{x}_{\text{DM}}(\omega)$
is then 
\begin{equation}
 |\tilde{x}_{\text{DM}}(\omega)|^2 = \frac{q_R^2}{m^2 \, L^2} \frac{1}{ (\omega^2 - \omega_c^2)^2 + \omega_c^4 \phi^2} \;.
 \label{eq:cxtDMSq}
\end{equation}
This result is identical to the result in eq.~\eqref{eq:rxtDMSq} taking into
account the exact matching in eqs.~\eqref{eq:fMatching}. This method is particularly
useful since KAGRA presents their experimental parameters in a form identical to eq.~\eqref{eq:cDEQ}.

There is also an important lesson, which we can draw from this equation
in comparison to eq.~\eqref{eq:cThPSD}.
Whenever we have a mode excited by a DM hit, the same mode will also
be excited by thermal energy. This was not taken properly into account
in the ThinET proposal in~\cite{Tsuchida:2019hhc} and we think that it would reduce
their SNR significantly.

\subsection{SNR}
\label{subsec:toymodelSNR}

To better quantify under which circumstances a potential DM signal
could be seen in a common earth-bound GW detector setup, we discuss
the signal-to-noise ratio (SNR), $\varrho^2$. Simply speaking this
is defined as the integral of signal over noise in the frequency
domain. Ideally we should also apply some kind of filter to enhance
the SNR. This is discussed in \cite{Moore:2014lga} and references therein.
The optimal SNR is given by
\begin{equation}
\varrho^2 = \int_{f_\text{min}}^{f_\text{max}} \diff f \frac{4 \, |\tilde{x}_{\text{DM}}(2 \, \pi \, f)|^{2}}{S_n(2 \, \pi \, f)} \;.
\label{eq:opSNR}
\end{equation}
In fact, in \cite{Moore:2014lga} the integration is chosen from zero to infinity.
In a realistic setup this is nevertheless not recommended, since at small
and large frequencies the background levels are very high suppressing
any signal component. We also did not take them into 
account in our toy setup.

Looking at Fig.~\ref{fig:ToyModel} it is suggestive to just integrate 
around the expected signal peak. The signal spectrum has the
shape of a Lorentzian with the peak frequency
$\omega_r (1 - \xi^2)$ and the full-width at half maximum (FWHM)
is at the frequencies
\begin{equation}
\label{eq:fwhmfre}
\begin{split}
\omega_{\text{max}}^2 &= \omega_r^2 (1 - \xi^2) + 2 \, \omega_r^2 \, \xi = \omega^{2}_{c} (1 + \phi)\;,\\
\omega_{\text{min}}^2 &= \omega_r^2 (1 - \xi^2) - 2 \, \omega_r^2 \, \xi = \omega^{2}_{c} (1 - \phi)\;,
\end{split}
\end{equation}
where we have used the exact result for $\omega_r$ and $\xi$
from eqs.~\eqref{eq:fMatching}.

Let us assume first that we have a situation at hand as in Fig.~\ref{fig:ToyModel},
where the quantum noise can be neglected compared to the thermal noise.
Then the SNR is
\begin{align}
\varrho_{\text{th}}^2 &=  \frac{1}{2 \pi} \int_{\omega_\text{min}}^{\omega_\text{max}} \diff \omega \frac{4 \, |\tilde{x}_{\text{DM}}(\omega)|^{2}}{S_n(\omega)}
 \approx 
 \frac{1}{4 \pi} \int_{\omega_\text{min}^2}^{\omega_\text{max}^2} \frac{\diff \omega^2}{\omega} \frac{4 \, |\tilde{x}_{\text{DM}}(\omega)|^{2}}{S_{\text{th}}(\omega)} \nonumber \\
 &=  \frac{q_R^2}{4  \, \pi \, m \, k_B \, T} \int_{\omega_\text{min}^2}^{\omega_\text{max}^2}  \frac{\diff \omega^2}{\omega_c^2 \, \phi} 
  =  \frac{1}{4 \pi} \frac{q_R^2}{m \, k_B \, T } \frac{ \omega_\text{max}^2 - \omega_\text{min}^2 }{\omega_c^2 \, \phi}
  = \frac{1}{4 \pi} \frac{E_R}{ E_{\text{th}} } \frac{ \omega_\text{max}^2 - \omega_\text{min}^2 }{\omega_c^2 \, \phi}  \nonumber \\
 &= \frac{1}{2 \pi} \frac{q_R^2}{m \, k_B \, T } = \frac{1}{2 \pi} \frac{E_R}{E_{\text{th}} } \;,
 \label{eq:SNRth}
\end{align}
where we have again used the exact result for $\omega_r$ 
and $\xi$
from eqs.~\eqref{eq:fMatching} and we have introduced the recoil energy
$E_R \equiv q_R^2/(2 \, m)$ and the thermal energy $E_{\text{th}} \equiv k_B \, T/2$
for one degree of freedom. The integration limits are based on eq.~\eqref{eq:fwhmfre}.
As one might have guessed naively, the SNR is proportional to the ratio of
DM scattering recoil energy and thermal energy.
For our example the SNR would be very small $\varrho^{2}_{\text{th}} = 4.09 \times 10^{-24}$. 
Mainly because the recoil energy is so tiny compared to the
thermal energy ($E_R = 3.37 \times 10^{-45}~\text{J}$ and $E_{\text{th}} = 1.31 \times 10^{-22}~\text{J}$). 
If we would follow the suggestion of
\cite{Cheng:2019vwy} instead and use $m = 10^{-6}$~kg with the other parameters kept the
same the SNR would be much larger, but still small
$\varrho^{2}_{\text{th}} = 9.33 \times 10^{-17}$.  
So we would need not only much lighter mirrors but also very low temperatures,
to enhance the SNR and make the signal actually observable. 

At this point we would like to give a disclaimer. The approximate formula presented here is good
for our toy model near the resonance. In the realistic case of KAGRA the ratio
$|\tilde{x}_{\text{DM}}|^2/(\omega \, S_{\text{th}})$ is a complicated function of $\omega$ and not
just a constant even near the peak. We will come back to this point later.

Let us now have a closer look at the quantum noise.
Actually, to have a peak in the DM signal can help to overcome the quantum noise
even though it is just in a narrow frequency window, cf.~Fig.~\ref{fig:ToyModel}.
In the experimentally difficult limit, where $T \approx 0$ and $S_\text{th} \ll S_{\text{qu}}$
even at the peak, we can approximate the SNR as 
\begin{align}
\varrho_{\text{qu}}^2 &=  \frac{1}{2 \pi} \int_{\omega_\text{min}}^{\omega_\text{max}} \diff \omega \frac{4 \, |\tilde{x}_{\text{DM}}(\omega)|^{2}}{S_n(\omega)} 
\approx \frac{1}{2 \pi} \int_{\omega_\text{min}}^{\omega_\text{max}} \diff \omega \frac{4 \, |\tilde{x}_{\text{DM}}(\omega)|^{2}}{S_{\text{qu}}(\omega)} \nonumber \\ 
 &\approx  \frac{q_R^2}{16 \, m \, \hbar \, \omega_c \, \phi } \left(1 + \frac{\pi - 8}{8 \, \pi} \phi^2 \right) = \frac{1}{16 \, \phi} \frac{E_R}{E_0}  \left(1 + \frac{\pi - 8}{8 \, \pi} \phi^2 \right) \;,
\label{eq:SNRqu}
\end{align}
where we have expanded in the small damping parameter $\phi$ and we have introduced
the zero energy of the quantum mechanical oscillator $E_0 \equiv \hbar \, \omega_c/2$.

In our example the peak of the DM signal actually overcomes the quantum noise
in a tiny frequency window, but we would still find a rather small SNR
$\varrho^{2}_{\text{qu}} = 5.74 \times 10^{-4}$.
However, with the hypothethical smaller target mass as suggested in \cite{Cheng:2019vwy}
where $m = 10^{-6}$~kg, 
the SNR value could be raised significantly to $\varrho^{2}_{\text{qu}} = 1.31 \times 10^{4}$ 
which is actually observable if we could suppress the suspension
thermal noise without suppressing the DM signal.

In summary, from this simple toy model we have already learned that conventional gravitational
wave detectors are not suited as detectors for light DM although they are extremely
sensitive. We have shown that the test 
mass $m$ and the temperature $T$ of the detector are two of the most crucial
properties affecting the sensitivity. Just reducing the test 
mass, the SNR increases since $\varrho^2 \sim 1/m$.
It is clear from eqs.~\eqref{eq:cThPSD}, 
\eqref{eq:sSQL} and \eqref{eq:cxtDMSq}, that the DM signal and 
noises are proportional to $1/m^2$ and $1/m$, 
respectively. Therefore, we can see there is a factor of $1/m$ 
enhancement of the SNR, see 
eqs.~\eqref{eq:SNRth} and \eqref{eq:SNRqu}. Besides, lower 
temperatures could further suppress the thermal noise. Hence, 
to make this idea work one would need ideally light 
and ultracold targets to increase the recoil energy and reduce 
the thermal noise.
Quantum sensing technologies which fulfill these criteria are currently being investigated
as DM detectors, for a recent overview see \cite{Carney:2020xol}.
In \cite{Monteiro:2020wcb} results for a DM search have been presented using
an optically levitated nanogram test mass cooled down to $\mathcal{O}(100)~\mu$K 
which is able to detect momentum transfers as low as 200~MeV/c but much higher than what we consider here. 
These experiments
are nevertheless rather new and small compared to KAGRA and other gravitational wave
detectors. Their noise is therefore still dominated by technical limitations and not
fundamental physics backgrounds as thermal or quantum noise. In the future that will
improve, but we refrain from providing a projection here, what their ultimate sensitivity
will be.

To reduce suspension thermal noise in gravitational wave detectors
there are actually proposals, 
i.e.~one where the noise is being filtered
out from the data by actively monitoring the suspension 
system~\cite{Santamore:2001nr}.
This idea is not being implemented so far as we know so we will not discuss it in great detail
here. Even if such a filter could distinguish between thermal noise and a potential DM signal
exploiting, for instance, cancellations in the spectrum as we discuss them in the next subsection,
one would still have to overcome the quantum noise, which is challenging in a conventional
GW detector setup. 

To distinguish a DM signal from thermal noise one could 
also change the temperature
and see if there is a noise component, which looks similar to 
thermal suspension noise but is not affected by the 
temperature. Based on eqs.~\eqref{eq:cThPSD} 
and \eqref{eq:cxtDMSq}, if the detector's temperature is 
lowered, the suspension thermal noise 
would be suppressed as mentioned before while a DM signal would 
remain unchanged.

\subsection{Cancellations in the DM signal spectrum}
\label{sec:Cancellation}

Before we apply what we learned from the above toy model to KAGRA,
we would like to extend our toy model a bit.

KAGRA and other earth-bound GW detectors consist
of a set of coupled damped harmonic oscillators. Therefore the motion
of the target mirrors cannot be described in terms of a single resonance
frequency. As the simplest generalization, we study
the case of a free motion with two eigenfrequencies and two damping factors, 
\begin{align}
x(t) = \theta(t) A \left( \text{exp}\left(- \omega_1 \, \xi_1 \, t \right) \sin (\omega_1 t) + r \,  \text{exp}\left(- \omega_2 \, \xi_2 \, t \right) \sin (\omega_2 t) \right) \;,
\end{align}
where $A$ and $r$ are real coefficients.
The frequencies $\omega_1$, $\omega_2$ 
and the damping factors $\xi_1$ and $\xi_2$
are all positive.
The Fourier transform squared in this case is, cf.~eq.~\eqref{eq:rxtDM},
\begin{align}
| \tilde{x}(\omega) |^2 &= A^2 \left|  \frac{1}{\omega_1^2 (1 + \xi_1^2) - \omega^2 + 2 \, \ci \omega \, \omega_1 \, \xi_1} + r  \frac{1}{\omega_2^2 (1 + \xi_2^2) - \omega^2 + 2 \, \ci \omega \, \omega_2 \, \xi_2}   \right|^2 \\
&\equiv A^2 \left| \hat{x}_1 + r \, \hat{x}_2 \right|^2 \;.
\label{eq:DoubleFT}
\end{align}

The two terms proportional to $|\hat{x}_1|^2$ and $|\hat{x}_2|^2$
just give two resonance-like peaks in the Fourier spectrum
similar to what we discussed before. On the other hand, for certain
parameters, the interference term
\begin{align}
\label{eq:Interference}
\begin{split}
\hat{x}_{12}^2 &\equiv 2 \,r\, \Re( \hat{x}_1 \hat{x}_2^\star ) \\
&= \frac{2 \, r \, [ (\omega^2 - \omega_1^2 (1 + \xi_1^2) ) ( \omega^2 - \omega_2^2 (1 + \xi_2^2) ) + 4 \, \omega^2 \omega_1 \omega_2 \xi_1 \xi_2   ] }{ ( (\omega - \omega_1)^2 + \omega_1^2 \xi_1^2 )((\omega + \omega_1)^2 + \omega_1^2 \xi_1^2 )( (\omega - \omega_2)^2 + \omega_2^2 \xi_2^2 )((\omega + \omega_2)^2 + \omega_2^2 \xi_2^2 )  } \;,
\end{split}  
\end{align}
can significantly suppress the total $| \tilde{x}(\omega) |^2$
for certain frequencies. In the case of driven oscillators 
this phenomenon is well known as anti-resonance.
In our previous toy setup, when we calculated the SNR, we focused 
on the frequency range around the resonance peaks. 
Therefore, our objective here is to investigate
under which conditions such an anti-resonance-like
behaviour can occur near a peak.

Inspired by the KAGRA parameters, we assume that $\xi_1$,~$\xi_2 \ll 1$
and $|\omega_i - \omega_j|/\omega_i \gg \xi_{i,j}$, where
$i,j = 1$,~2. Based on these assumptions and looking at the above expression
for $\hat{x}_{12}^2$ we have to consider the interference term
at $\omega \approx \omega_1$ or $\omega \approx \omega_2$.
Since the problem is basically symmetric under exchange of the labels, it is sufficient to consider only one case. Taking the latter case as an example
\begin{align}
\hat{x}_{12}^2(\omega_2)
&=
\frac{2 \, r \, [ - \xi_2 (\omega_2^2 - \omega_1^2 (1 + \xi_1^2) ) + 4 \, \omega_1 \omega_2 \xi_1    ]}
{\omega_2^2 \xi_2 (4 + \xi_2^2) ( (\omega_2 - \omega_1)^2 + \omega_1^2 \xi_1^2 )((\omega_2 + \omega_1)^2 + \omega_1^2 \xi_1^2 )} \label{eq:sInterference}\\
&\approx
\frac{r \, [ - \xi_2 (\omega_2^2 - \omega_1^2 ) + 4 \, \omega_1 \omega_2 \xi_1    ]}
{2 \, \omega_2^2 \xi_2  (\omega_2^2 - \omega_1^2)^2 }\;.
\end{align}
The denominator of eq.~\eqref{eq:sInterference} is always
positive and a cancellation with the other (positive) terms in $|\tilde{x}(\omega)|^2$ can only occur
if the nominator is negative, i.e.~if $r \, [ - \xi_2 (\omega_2^2 - \omega_1^2 (1 + \xi_1^2) ) + 4 \, \omega_1 \omega_2 \xi_1    ] < 0$.
Furthermore, to be noticeable the interference term should be similar
in size as the other two terms. Let us assume that we can neglect
the contribution from the first resonance compared to the second
term for simplicity, i.e.
\begin{align}
|\hat{x}_1(\omega_2)|^2 \ll r^2 |\hat{x}_2(\omega_2)|^2 =
\frac{r^2}{  \omega_2^4 \xi_2^2 (4 + \xi_2^2) } \;.
\end{align}
Therefore, under these assumptions a cancellation occurs if
$\hat{x}_{12}^2(\omega_2) \approx - r^2 |\hat{x}_2(\omega_2)|^2$,
i.e.
\begin{align}
\label{eq:AntiresonanceCond}
\begin{split}
 - \frac{r}{2 } &\approx  
\frac{ - \omega_2^2 \xi_2 (\omega_2^2 - \omega_1^2 (1 + \xi_1^2) ) + 4 \, \omega_1 \omega_2^3 \xi_1 }
{ ( (\omega_2 - \omega_1)^2 + \omega_1^2 \xi_1^2 )((\omega_2 + \omega_1)^2 + \omega_1^2 \xi_1^2 )} \\
&\approx
\frac{-\omega_2^2 \xi_2}{( \omega_1^2 - \omega_2^2 )} + \frac{4 \, \omega_1 \, \omega_2^3 \, \xi_1}{( \omega_1^2 - \omega_2^2 )^2} \;.
\end{split}
\end{align}
That means that $r$ must be small of the order of $\mathcal{O}(\xi_{i,j})$. If we would
look at the other case, where a cancellation occurs near
$\omega \approx \omega_1$ then $1/r$ must be small, $\mathcal{O}(\xi_{i,j})$.

For the calculation of the SNR we also assume that we only need to take into
account the thermal noise from the second mode and we neglect quantum noise
\begin{align}
 S_n \approx S_{\text{th},2}(\omega) = \frac{4\, k_{B} \, T}{L^{2}} \, \frac{2\, \omega_{2}\, \xi_{2}/m}{(\omega^{2}-\omega^{2}_{2}(1-\xi^{2}_{2}))^{2}+4\, \omega^{4}_{2}\, \xi^{2}_{2}} \;,
\label{eq:Sth2} 
\end{align}
which is consistent with our previous assumptions.
We also use the FWHM boundaries for the integration as if there
was no cancellation or second resonance peak.
The peak we consider here is at $\omega_2^2 (1 - \xi_2^2)$
and the FWHM is located at $\pm 2 \, \omega_2^2 \xi_2$ around the peak.
The interference term has an extremum at $\omega_2$ meaning that it is
situated within this FWHM range.

The SNR is then given by 
\begin{align}
\varrho^2 &\approx  \frac{2 \, A^2}{\pi} \int_{\omega_\text{min}}^{\omega_\text{max}} \diff \omega \frac{ r^2 \, |\hat{x}_{2}(\omega)|^{2} +  \hat{x}_{12}^2(\omega) }{S_{\text{th},2}(\omega)} \nonumber\\ 
&= \frac{A^2 r^2  m L^2}{2 \, \pi \, k_B \, T} \left(1- \frac{4 \,\xi_2^2}{3} \frac{\omega_2^2 (5 \,  \omega_2^2 - 3 \, \omega_1^2) }{ r\, (\omega_2^2 - \omega_1^2)^2 }\right) + \mathcal{O}(\xi_2^4,\xi_1)\;. 
\end{align}
Thus, given that $\xi_2 \ll 1$ we expect the correction to the SNR
to be very small. Even in the case of a strong cancellation, 
i.e.~$\hat{x}_{12}^2(\omega_2) \approx - r^2 |\hat{x}_2(\omega_2)|^2$,
this would still be true since then $r$ is $\mathcal{O}(\xi_1,\xi_2)$,
cf.~eq.~\eqref{eq:AntiresonanceCond}.
Based on this, we conclude that the occurrence of cancellations in the signal spectrum
does not significantly reduce the expected SNR.

At this point, we already want to point out, that these cancellations in the
spectrum are characteristic for the DM noise, but not the thermal suspension noise.
We will see this very clearly, when we compare a potential DM signal with
the noise spectrum in KAGRA. The DM signal features many prominent dips while the noise
does not. In a dedicated DM experiment these differences could be optimised to allow
a better distinction of signal and noises.

\section{A realistic example: KAGRA}
\label{sec:KAGRA}

After having discussed some essential features using a simple toy model,
we now want to turn to the concrete example of KAGRA \cite{KAGRAnoise}.
The results for other earth-bound GW detectors based on
laser interferometers, like LIGO or VIRGO, would be qualitatively very similar.
Due to the cryogenic system KAGRA is actually better for our purposes since
thermal noise is smaller then.

Let us briefly provide some essential details about the experimental
setup in KAGRA. KAGRA consists of a double pendulum connected by blade springs (BSs).
The first pendulum is represented by the suspended intermediate mass (IM)
connected by 4 CuBe wires to the upper part of the system. The
second pendulum corresponds to the TM suspended by 4 sapphire wires attached to the BSs.
Mathematically this corresponds to a triple oscillator system. The three
kinds of masses relevant here are the IM, $m_1$,
the BS,
$m_2$, and the TM, $m_3$, where we closely follow
the notation and labelling as in \cite{KAGRAnoise} with 
only minor
modifications. The relative positions of the four TMs are probed
by the interferometer system. We will also use $i=1$, $2$, $3$ for the
relative coordinates of the three kinds of masses. To be precise, we are
only concerned with the vertical (horizontal)
dimensionless, relative displacements, which we label as
$x_{iv}(t)$ ($x_{ih}(t)$), $i=1$, $2$, $3$. These two sets of coordinates
are orthogonal to each other and can be treated independently.
It should be noted that in Komori's code from~\cite{KAGRAnoise}
the displacements are not dimensionless which leads to some
differences in the appearance of factors of $L$.

We split our discussion again into two parts. First, we discuss the noise
which is completely based on~\cite{KAGRAnoise} and we
use the detuned resonant side-band extraction (DRSE) 
configuration to be precise.
We then describe DM scattering in KAGRA and compare it to the noise
and also briefly comment on the SNR.

\subsection{Noise}
\label{sec:KAGRAnoise}

Terrestrial GW detectors generically
suffer from seismic noise at low frequencies below
10~Hz and at large frequencies above 1~kHz other sources
of noise dominate. For KAGRA in this region actually quantum
noise gets dominant, cf., for instance, \cite{KAGRAnoise,Buonanno:2001cj}.
In the region in between,
the KAGRA noise budget is mostly dominated by suspension thermal noise
and quantum noise. For that reason we have focused on these
two sources of noise in our toy model and we also focus on 
these two here.

\subsubsection{Vertical Suspension Thermal Noise}
\label{subsec:VTh}

We begin our discussion of the thermal noise with the vertical direction
which is easier compared to the horizontal one as we will see later.
As we mentioned already the KAGRA detector can be modelled
as three coupled (damped) harmonic oscillators
in terms of the coupled differential equation \cite{KAGRAnoise}
\begin{align}
\left( M \frac{\diff^{2}} {\diff t^{2}} + K_{v} \right) \vec{x}_{v}(t) = \frac{\vec{F}_{\text{ext},v}(t)}{L} ,
\label{eq:vtheom1}
\end{align} 
where 
\begin{align}
\label{eq:mkv}
M =\begin{pmatrix}
m_{1} & 0
& 0  \\
0
& 
m_{2} 
&  0 \\
0  & 0 &  m_{3}  
\end{pmatrix}  \;&, \quad
K_{v} = \begin{pmatrix}
K_{1v}+K_{2v} & -K_{2v}
& 0  \\
-K_{2v}
& 
K_{2v}+K_{3v} 
&  -K_{3v} \\
0  & -K_{3v} &  K_{3v}  
\end{pmatrix} \;, \\
\vec{x}_{v}(t) = \begin{pmatrix}
x_{1v}(t) \\
x_{2v}(t) \\
x_{3v}(t)
\end{pmatrix} &\text{ and }
\vec{F}_{\text{ext},v}(t) = \begin{pmatrix}
F_{\text{ext},1v}(t) \\
F_{\text{ext},2v}(t) \\
F_{\text{ext},3v}(t)
\end{pmatrix} \;.
\label{eq:xF}
\end{align}
The spring constants $K_{iv} \equiv k_{iv}(1+\text{i} \, 
\phi_{iv})$ are complex. As we have discussed in our toy
model the imaginary parts damp the oscillations.
The numerical values of the 
parameters in eq.~\eqref{eq:mkv} are for the convenience
of the reader taken from \cite{KAGRAnoise} and  collected in 
Tab.~\ref{tab:Vparams}.

Here, we can see the advantage of the complex spring constant
notation compared to a real notation. To calculate the thermal noise
we assume an external periodic force and it is not immediately clear
which frequencies we would have to use in the matrix generalisation
of the damping term in eq.~\eqref{eq:rDEQ}. For the complex spring
constants this question does not arise. All the information is 
contained in $K_v$ and $M$.

\begin{table}
	\centering
	\begin{tabular}{lr}
		\toprule
		Parameters & Values\\
		\midrule
		$m_{1}$   & 20.5 kg \\
		$m_{2}$ & 0.22 kg \\ 
		$m_{3}$ & 22.8 kg \\
		\midrule
		$k_{1v}$ & 5.63 $\times 10^{5}$ N/m \\
		$k_{2v}$ & 1.91 $\times 10^{5}$ N/m \\
		$k_{3v}$ & 9.19 $\times 10^{6}$ N/m\\
		\midrule
		$\phi_{1v}$ & 4.21 $\times 10^{-6}$ \\
		$\phi_{2v}$ & 5.89 $\times 10^{-7}$ \\
		$\phi_{3v}$ & 2.0 $\times 10^{-7}$ \\
		\bottomrule
	\end{tabular}
	\caption{
		The values of masses and vertical, complex spring constants in KAGRA
		appearing in eq.~\eqref{eq:mkv} taken from \cite{KAGRAnoise}.
		See also main text for more details.} 
	\label{tab:Vparams}
\end{table}

We can now calculate the thermal noise in a very similar way to what
we discussed before in Section~\ref{subsec:toymodelthermalnoise}.
We only have to change everything into vector and matrix quantities.
First of all, we calculate the admittance by assuming again that the
system is subject to a periodic external force,
i.e.~$F_{\text{ext}, iv}(t)=\tilde{F}_{\text{th},iv}(\omega) 
\exp(\ci \omega t)$ and the corresponding displacement  
$x_{iv}(t)=\tilde{x}_{iv}(\omega) \exp(\ci \omega t)$ related to it by
\begin{align}
D_{v}(\omega) \, \vec{\tilde{x}}_{v}  \equiv \left( -\omega^{2} \, M + K_{v} \right) \vec{\tilde{x}}_{v} = \frac{\vec{\tilde{F}}_{\text{th},v}}{L} \;.
\label{eq:vtheom2}
\end{align}
The corresponding $D_c(\omega)$ in the toy model, 
cf.~eq.~\eqref{eq:YcEQ}, was a number while here
it is a matrix. And as we have seen before the admittance $Y_v$ 
can be directly related to
$D_v(\omega)$ via
\begin{align}
Y_{v}(\omega) \equiv \text{i} \, \omega \, D_{v}^{-1}(\omega)\;.
\label{eq:YvD}
\end{align}
Now we can again use the fluctuation-dissipation theorem \cite{Callen:1951,Callen:1952} to relate
the admittance to the thermal noise spectrum which is also 
a matrix here
\begin{align}
S_{x_{iv}x_{jv}}(\omega) = \frac{4 k_{B} \, T \, \Re(Y_v(\omega))_{ij}}{L^{2} \,  \omega^{2}}\;.
\label{eq:FDT1}
\end{align}
The diagonal elements of this matrix represent the thermal
noise PSD and since we are probing the TM, the third
coordinate, we are interested in
\begin{align}
S_{\text{th},v}(\omega) \equiv S_{x_{3v} x_{3v}}(\omega) = \frac{4 \, k_{B} 
	\, T_{3} \, \Re(Y_v)_{33} }{L^{2} \, \omega^{2}}
\; , 
\label{eq:Svth}
\end{align}
where $T_{3}=19$~K (the three masses actually have slightly different temperatures).
With the definition 
in eq.~\eqref{eq:strainthermal}, the strain amplitude for the vertical 
thermal noise can be readily obtained as 
\begin{align}
h_{\text{th},v}(\omega) =\text{VHC} \, \sqrt{4|S_{\text{th},v}(\omega)|} .
\label{hvth}
\end{align}
where the factor of four refers to the fact that there are four equal TMs. The 
VHC factor accounts for the fact that there is a non-vanishing component of the 
horizontal motion, which is being probed by the experiment 
due to the tilt of the interferometer baseline. The 
value of this factor is $1/200$ \cite{KAGRAnoise}.

\begin{figure}
	\centering
	\includegraphics[width=0.8\textwidth]{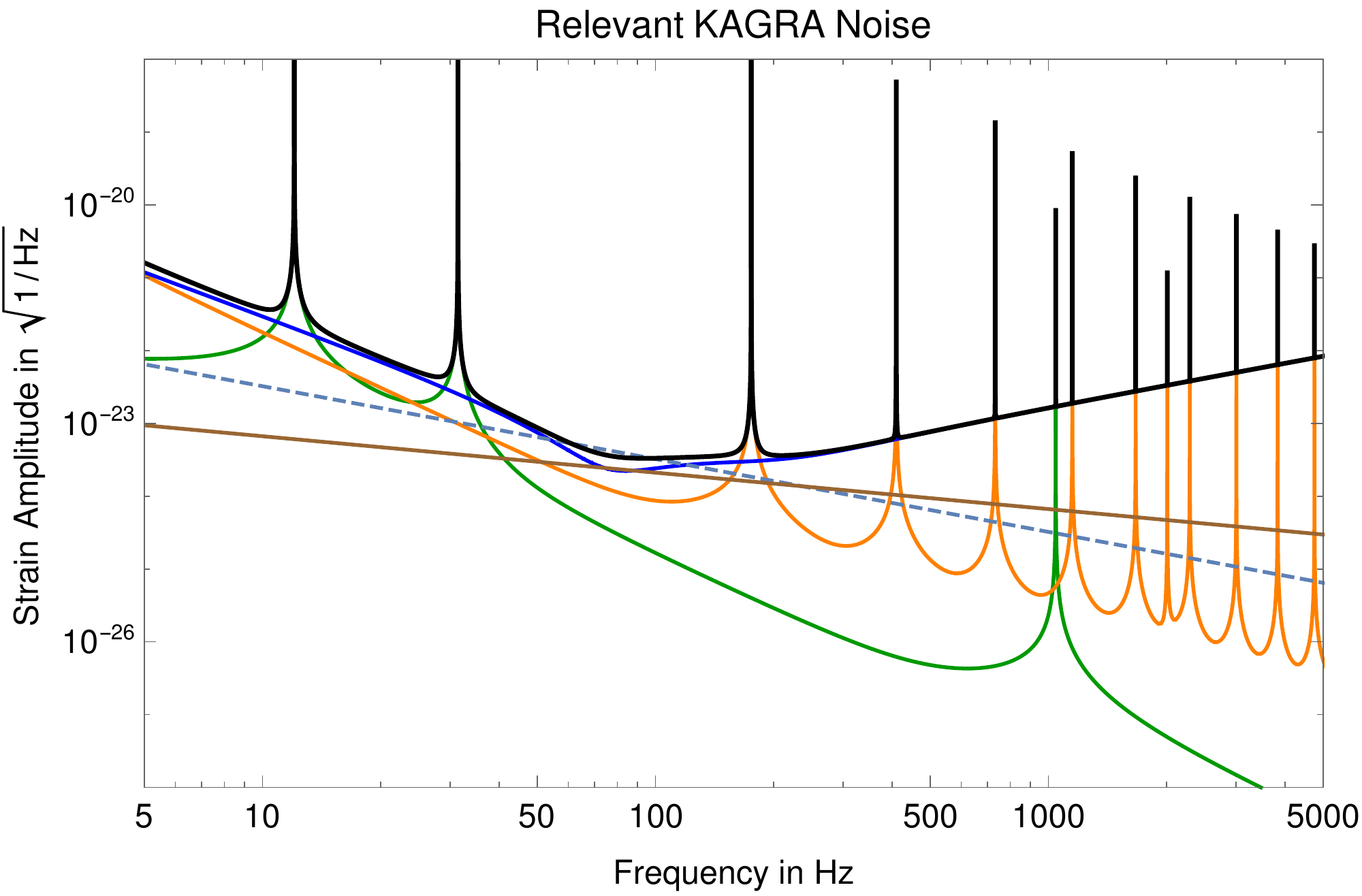}
	\caption{Overview of strain amplitudes of relevant noise components in KAGRA.
		The green (orange) line is the vertical (horizontal) suspension thermal
		noise. The KAGRA quantum noise is shown as blue line and the SQL is 
		shown for comparison as light blue, dashed line. The mirror thermal noise is the brown line. The total noise is the black line.
        } 
	\label{fig:KAGRAnoises}
\end{figure}

We show the strain of the vertical thermal suspension noise
in Fig.~\ref{fig:KAGRAnoises} as the green line. We can clearly
identify three peaks which is what we expect for a system
of three coupled harmonic oscillators which exhibits three
normal modes. The respective peak frequencies are 12.04, 31.43
and 1044.09 Hz. We would like to mention that compared to the
figures in many KAGRA documents, for instance in \cite{KAGRAnoise},
our peaks for the noise look much higher. That is due to the difficulty
of plotting very narrow resonances. We checked, that we agree with
Komori's code in~\cite{KAGRAnoise} after increasing their plot resolution drastically.

\subsubsection{Horizontal Suspension Thermal Noise}
\label{subsec:HTh}

We now turn to the horizontal thermal noise which - although very similar -
still has some complications compared to the vertical case.
We can again write
\begin{align}
\left( M \frac{\diff^{2}} {\diff t^{2}} + K_{h} \right)\vec{x}_{h}(t) = \frac{\vec{F}_{\text{ext},h}(t)}{L} ,
\label{eq:htheom1}
\end{align} 
with $M$ as before and $K_h$ has the same structure as $K_v$.
Apart from different numerical values for $K_{2h}$
compared to $K_{2v}$ there are two differences.
First, $K_{1h}$ is an effective spring constant given by
\begin{align}
K_{1h} = \frac{4 \, \tau_1}{l_1 - 2\sqrt{\frac{E_{1h} \, I_1}{\tau_1}}} \equiv k_{1h} (1 + \ci \phi_{1h}) \;,
\label{eq:k1h}
\end{align}
where $\tau_1 = (m_1 +m_2 + m_3) \, g /4$ with g the gravitational 
acceleration, $E_{1h}$ and $l_1$ are the 
associated tension per wire, complex Young modulus of the wire and 
the length of the wire connecting to IM.

\begin{table}
	\centering
	\begin{tabular}{lr}
		\toprule
		Parameters & Values\\
		\midrule
		$k_{1h}$ & 1.67 $\times 10^{3}$ N/m \\
		$k_{2h}$ & 3.47 $\times 10^{7}$ N/m \\
		\midrule
		$\phi_{1h}$ & 4.59 $\times 10^{-8}$ \\
		$\phi_{2h}$ & 5.89 $\times 10^{-7}$ \\
		\midrule
		$\Re(E_{w})$ & 4.0 $\times 10^{11}$ Pa \\
		$\Im(E_{w})$ & 8.0 $\times 10^{4}$ Pa \\
		$I_{cw}$ & 3.22 $\times 10^{-13}$ $\text{m}^{4}$ \\
		$\rho_{l}$ & 8.04 $\times 10^{-3}$ $\text{kg/m}$ \\
		$l_{1}$ & 0.26 m \\
		$l_{2}$ & 0.35 m \\
		$g$ & 9.81 $\text{m}/ \text{s}^{2}$ \\
		\bottomrule
	\end{tabular}
	\caption{
		The values of the horizontal, complex spring constants in KAGRA
		and related material parameters appearing implicitly in eq.~\eqref{eq:htheom1} taken from \cite{KAGRAnoise}.
		See also main text for more details.} 
	\label{tab:Hparams}
\end{table}

Second, the horizontal spring constant for the TM
is frequency dependent, 
i.e.~$K_{3h} = K_{3h}(\omega)$.
The analytical expression for $K_{3h}(\omega)$ is written in
terms of the frequency dependent wave number of the elastic
wire, $k_{s}$, and the
flexural stiffness of the  wire, $k_{e}$, defined as
\begin{align}
k_{s,e}^2(\omega) = \frac{\mp \tau_3 + \sqrt{\tau^{2}_3+4 \, E_{w} \,  I_{cw} \, \rho_{l} \, \omega^{2}}}{2 \, E_{w} \, I_{cw}} \; ,
\label{eq:kse}
\end{align}
where $\tau_3  = (m_{3} \, g)/4$, $E_w$, $I_{cw}$, and $\rho_{l}$ are the
associated tension 
per wire, complex Young modulus of the wire, area moment of inertia of
the wire, and linear mass density of the wire respectively. The tension per wire is
simply obtained by dividing the total tension due to the weight of $m_{3}$ 
by the number of wires (which is four in this case).
Using these definitions, $K_{3h}(\omega)$ can be written as \cite{KAGRAnoise}
\begin{align}
K_{3h}(\omega) = 4 \, E_{w} \, I_{cw} \, k_{e} \, k_{s} \, \left(k^{2}_{e}+k^{2}_{s}\right) R_{3h}(\omega) \;,
\label{eq:k3h0}
\end{align}
where the factor $R_{3h}(\omega)$ is given by
\begin{equation}
R_{3h}(\omega) = \frac{k_{e} \, \cos(k_{s} \, l_{2}) \, \sinh(k_{e} \, l_{2}) + k_{s} \, \sin(k_{s} \, l_{2}) \, \cosh(k_{e} \, l_{2})}{2 \, k_{e} \, k_{s} \left[1-\cos(k_{s} \, l_{2}) \, \cosh(k_{e} \, l_{2})\right] + (k^{2}_{e} - k^{2}_{s}) \, \sin(k_{s} \, l_{2}) \, \sinh(k_{e} \, l_{2})}  \;, 
\label{eq:F3h}
\end{equation}
where $l_{2}$ is the length of the TM 
suspension wire.
The numerical values of the parameters for the 
horizontal equation of motion, eq.~\eqref{eq:htheom1}, are collected
for the convenience of the reader in Tab.~\ref{tab:Hparams}.

By repeating the same steps as before for the vertical noise in
eqs.~\eqref{eq:vtheom2}-\eqref{eq:Svth}, we get the strain amplitude of
horizontal 
thermal noise as
\begin{align}
h_{\text{th},h}(\omega) = \sqrt{4|S_{\text{th},h}(\omega)|}  \;.
\label{eq:hhth}
\end{align}
where $S_{\text{th},h}(\omega) \equiv S_{x_{h3}x_{h3}}(\omega)$ analogous to 
eq.~\eqref{eq:FDT1}. Since we now discuss the horizontal modes directly
we do not need to include the VHC factor.

We show the strain of the horizontal thermal suspension noise
in Fig.~\ref{fig:KAGRAnoises} as the orange line. One immediate
difference of the horizontal compared to the vertical noise
is the much larger number of peaks in the horizontal noise.
This is due to the $\omega$ dependence of $K_{3h}$ which leads
to so-called violin modes. As the name suggests they originate
from vibrational excitations of the suspension wires being
thermally excited, see~\cite{Gonzales:1994}.
There are a total of ten
modes in the window between 5~Hz and 5~kHz so we will not list them
all here explicitly. The first peak at 175.44 Hz
corresponds approximately to the toy model example though.

\subsubsection{Other noise components}
\label{subsec:OtherNoise}

The KAGRA quantum noise, $S_{\text{qu}}$, is actually
not correctly described by the SQL, cf.~eq.~\eqref{eq:SQL}. KAGRA is 
a second-generation interferometer with a signal recycling cavity
where the quantum noise is more complicated. 
Since we focus here on DM and the explicit, complicated expressions for the
quantum noise do not give us 
any insights on that, we just show the curves for illustration in 
Fig.~\ref{fig:KAGRAnoises}. For more details,
see~\cite{Buonanno:2001cj,KAGRAnoise}. 

Between about 75 and 88 Hz, the mirror
thermal noise, $S_{\text{mir}}$, which is the sum of
thermoelastic noise~\cite{Somiya:2010}, 
substrate thermal noise~\cite{Somiya:2011np} and coating thermal 
noise~\cite{Harry:2001iw}, is actually dominant over 
the suspension thermal noise and quantum noise. However, since 
this noise component does not contribute significantly to the total 
noise around the resonance peaks, we again just show the curve of this noise in 
Fig.~\ref{fig:KAGRAnoises}.
Since the mirror thermal noise is also proportional to the mirror temperature
like the suspension thermal noise it could be suppressed 
by reducing the temperature as well.

The total noise we consider is given by
\begin{align}
S_{\text{tot}}(\omega) = 4 \, \text{VHC}^{2} \,S_{\text{th},v}(\omega)+ 4 \, S_{\text{th},h}(\omega) + S_{\text{qu}}(\omega) + 2 \, S_{\text{mir}}(\omega) \;,
\label{eq:Stot}
\end{align}
where the factors of four are due to having four TMs whereas
there is only a factor of two for the mirror thermal noise as there
are two different coatings used. The quantum noise already includes
any such factors.
The strain is then given by
\begin{align}
h_{\text{tot}}(\omega) = \sqrt{h^{2}_{\text{th},v}(\omega)+h^{2}_{\text{th},h}(\omega) + h^2_{\text{qu}}(\omega) + h^2_{\text{mir}}(\omega)} \;.
\label{eq:htot}
\end{align}

Within the frequency range we focus on, all the noise 
components are shown in Fig.~\ref{fig:KAGRAnoises} for 
illustration.
We show the strain of the quantum noise as blue line
and the mirror noise as brown line
in Fig.~\ref{fig:KAGRAnoises}. For comparison we also show the 
SQL as dashed blue line. The actual quantum noise is smaller than 
the SQL in the window between 54~Hz and 120~Hz
where KAGRA is most sensitive. In the region where the resonance
modes of the suspension system appear the SQL is instead better than
the actual quantum noise. For KAGRA this makes sense since they
do not want to have any thermally excited modes in their most sensitive
region. For our purposes that should be constructed differently and
the quantum noise should be minimal near a resonance where the DM signal
can be comparatively large.

We also show in Fig.~\ref{fig:KAGRAnoises} the combination of all
relevant noise components as black line.
For the first three peaks of 
the total noise, the strain amplitudes can reach between
$\mathcal{O}(10^{-16})$ and $\mathcal{O}(10^{-18})$~Hz$^{-1/2}$.
We will use this combination in our later figures in
comparison to a potential DM signal.

\subsection{DM signal}

We now turn to the modelling of a hypothetical DM signal,
which we do in analogy to our discussions for the toy model.
Again we want to assume that the DM hit occurs at $t = 0$
and it generally transmits a recoil momentum in horizontal
and vertical direction.

Let us first discuss the effect of a hit in the TM in the 
vertical direction first.
In this case, the force can be described by
\begin{equation}
F_{\text{DM},3v}(t) = q_{R,v} \, \delta(t) \;,
\label{eq:F3vDM}
\end{equation}  
where $q_{R,v}$ is the recoil momentum of DM in vertical
direction. The equation of motion for this case is given by eq.~\eqref{eq:xF}
with $F_{\text{ext},3v}$ replaced by $q_{R,v} \, \delta(t)$.
Taking the Fourier transform of it we immediately arrive at
\begin{align}
\vec{\tilde{x}}_{v}(\omega) = D^{-1}_{v}(\omega)  \frac{\vec{\tilde{F}}_{\text{DM},v}(\omega)}{L}
= D^{-1}_{v}(\omega) \begin{pmatrix}
0 \\
0 \\
\frac{q_{R,v}}{L}
\end{pmatrix} \;,
\label{eq:DDM}
\end{align}
or in component notation
\begin{align}
\tilde{x}_{iv}(\omega) = \sum_{j=1}^{3}  \left(D^{-1}_{v}(\omega) \right)_{ij} \frac{\tilde{F}_{j,v}(\omega)}{L} \;.
\label{eq:invDFDM}
\end{align}
Since we are probing the third component and the force is assumed
to be in the third component as well we get easily
\begin{equation}
|\tilde{x}_{\text{DM},v}(\omega)|^2 \equiv |\tilde{x}_{3v}(\omega)|^2 =  
\left| \left(D^{-1}_{v}(\omega) \right)_{33}  \frac{\tilde{F}_{3,v}(\omega)}{L} \right|^2 \;,
\label{eq:vTMDM}
\end{equation}
where $\tilde{F}_{3,v}(\omega) = q_{R,v}$.
For a hit in the IM the formula would be instead
\begin{equation}
|\tilde{x}_{\text{DM},v}(\omega)|^2 =  
\left| \left(D^{-1}_{v}(\omega) \right)_{31}  \frac{\tilde{F}_{1,v}(\omega)}{L} \right|^2 \;,
\label{eq:vIMDM}
\end{equation}
with $\tilde{F}_{1,v}(\omega) = q_{R,v}$. The generalisation
to a hit in the BS or observation of other components is 
straightforward.

The corresponding strain amplitude induced by such a DM hit
taking the tilt of the interferometer baseline into account is
\begin{align}
h_{\text{DM},v}(\omega) = \text{VHC} \, \sqrt{\frac{2 \, \omega}{\pi}  |\tilde{x}_{\text{DM},v}(\omega)|^2 } \;,
\label{eq:hvDM}
\end{align}
where again $\text{VHC} = 1/200$.

The DM induced strain in horizontal direction can be derived completely
analogous to the vertical case
\begin{align}
h_{\text{DM},h}(\omega) = \, \sqrt{\frac{2 \, \omega}{\pi}  |\tilde{x}_{\text{DM},h}(\omega)|^2 } \;,
\label{eq:hhDM}
\end{align}
where, i.e.
\begin{equation}
|\tilde{x}_{\text{DM},h}(\omega)|^2 \equiv |\tilde{x}_{3h}(\omega)|^2 =  
\left| \left(D^{-1}_{h}(\omega) \right)_{33}  \frac{\tilde{F}_{3,h}(\omega)}{L} \right|^2 \;,
\label{eq:hTMDM}
\end{equation}
for a horizontal hit in a TM with $\tilde{F}_{3,h}(\omega) = q_{R,h}$. Again the 
generalisation to other components is straightforward.

For the total DM signal we use
\begin{equation}
	|\tilde{x}_{\text{DM}}(\omega)|^2 =	\text{VHC}^2 |\tilde{x}_{\text{DM},v}(\omega)|^2	 + |\tilde{x}_{\text{DM},h}(\omega)|^2 \;,
\end{equation}
where we have included the VHC$^2$ factor for the vertical hits since
we only observe this through the tilt in the setup.

\begin{figure}
	\centering
	\includegraphics[width=0.8\textwidth]{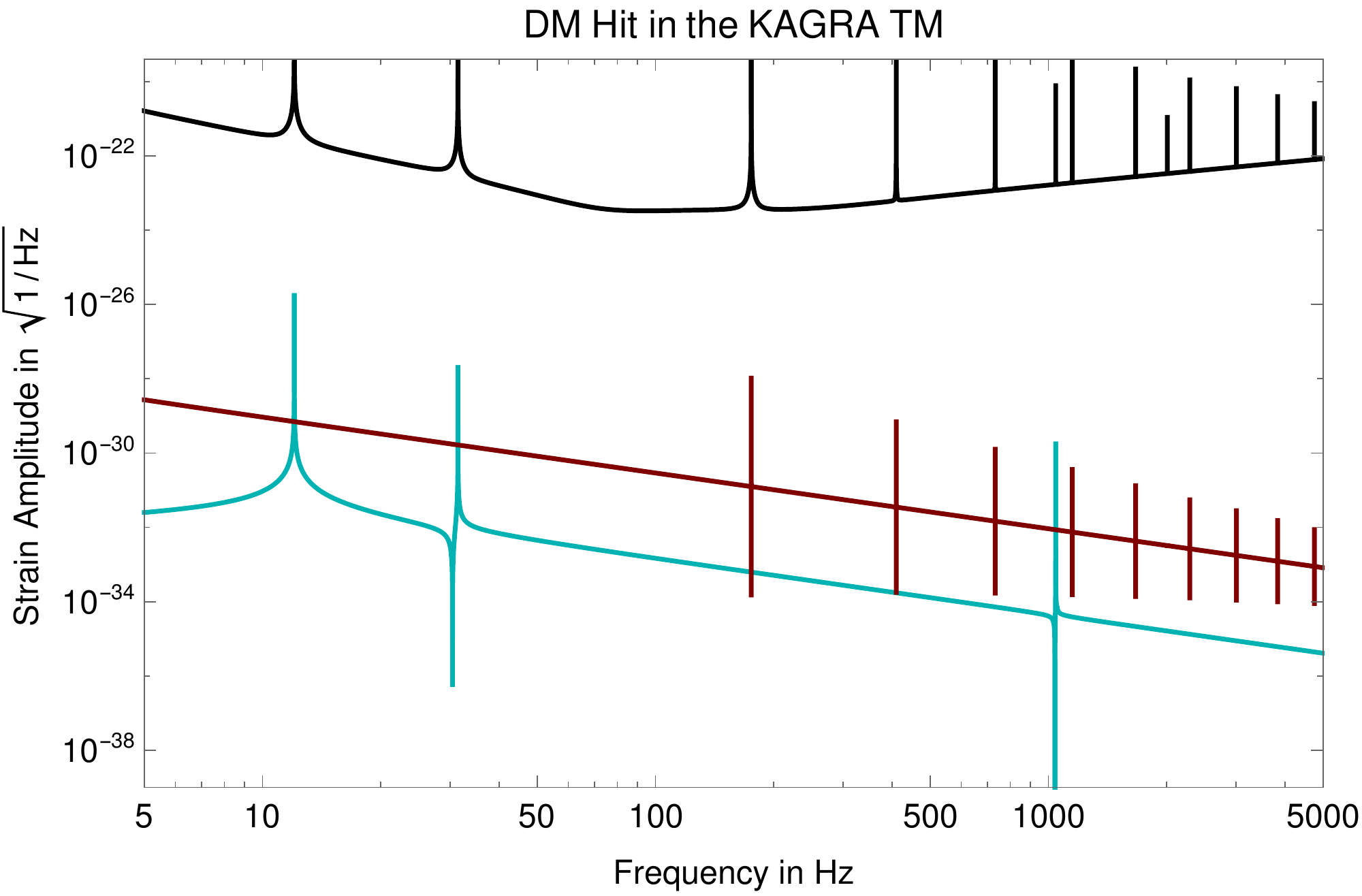}
	\caption{
		Comparison of strain amplitudes of noise shown as
		black line and a DM hit in the TM in vertical (horizontal) direction
		shown as aqua (maroon) line. For more details, see main text.
	}
	\label{fig:TMKagra}
\end{figure}

\begin{figure}
	\centering
	\includegraphics[width=0.8\textwidth]{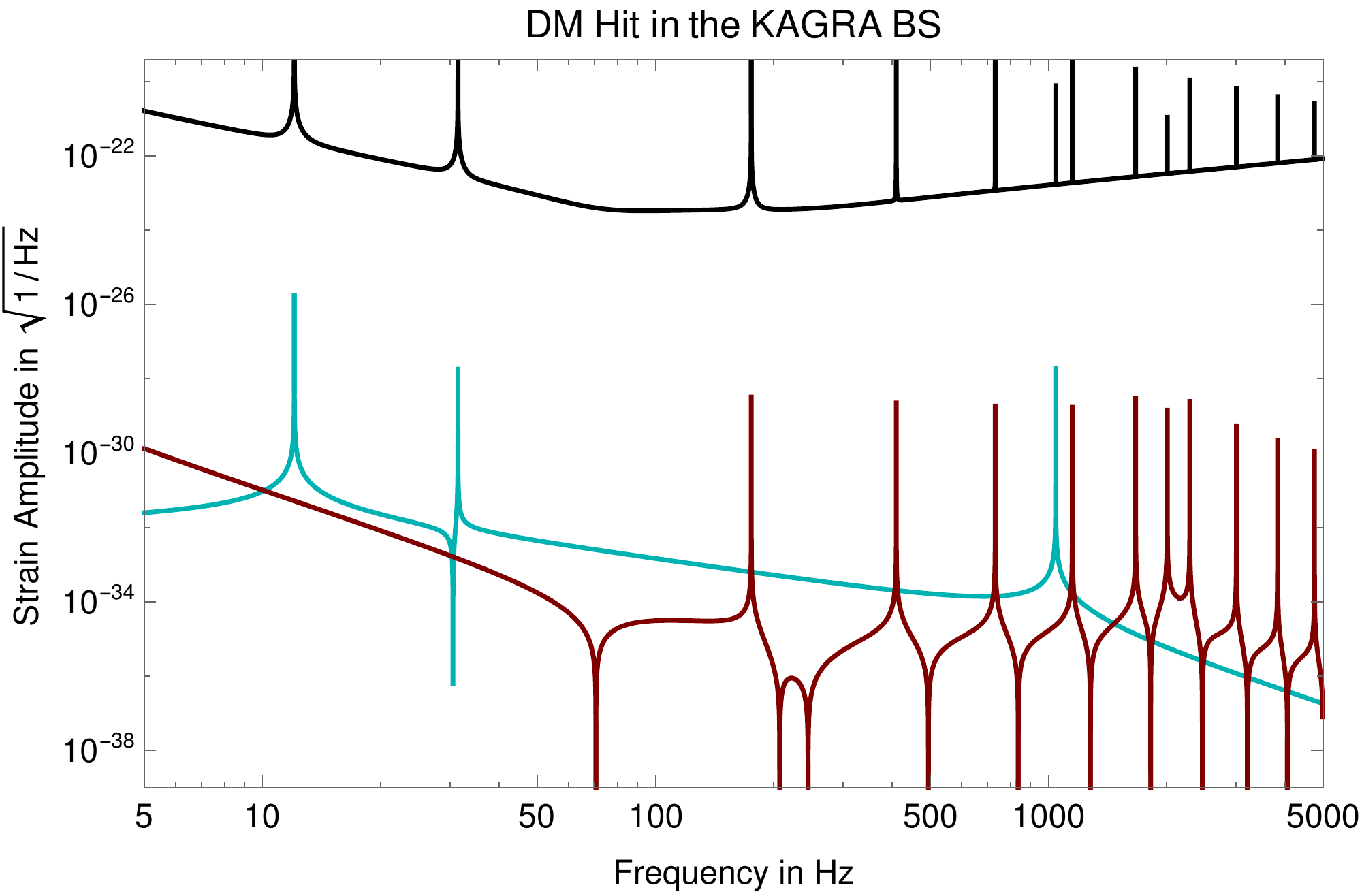}
	\caption{
		Comparison of strain amplitudes of noise shown as
		black line and a DM hit in the BS in vertical (horizontal) direction
		shown as aqua (maroon) line. For more details, see main text.
		}
	\label{fig:BSKagra}
\end{figure}

\begin{figure}
	\centering
	\includegraphics[width=0.8\textwidth]{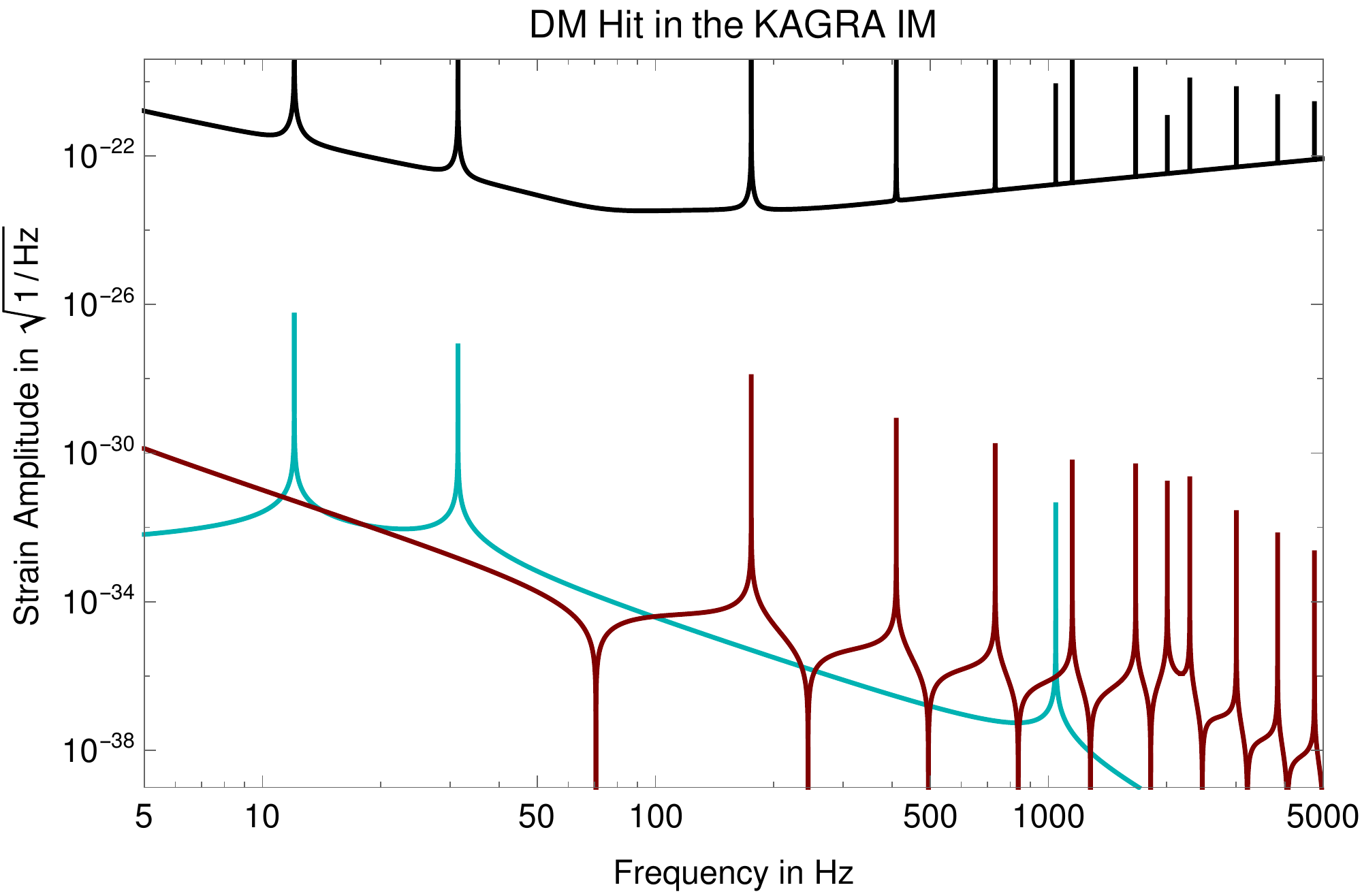}
	\caption{
		Comparison of strain amplitudes of noise shown as
		black line and a DM hit in the IM in vertical (horizontal) direction
		shown as aqua (maroon) line. For more details, see main text.}		
	\label{fig:IMKagra}
\end{figure}

In Figs.~\ref{fig:TMKagra}, \ref{fig:BSKagra} and \ref{fig:IMKagra}
we show a comparison of DM hits in the TM, BS and IM respectively.
In these figures, we show two DM cases each, assuming that 
our estimate for the recoil momentum,
$q_{R} \sim m_{\text{DM}} |\vec{v}_{\text{DM}}|$ with the DM
mass, $m_{\text{DM}}= 1$~GeV/$\text{c}^2$ and the typical 
velocity, $|\vec{v}_{\text{DM}}|=220$~km/s is either 
completely in vertical direction (aqua line) or horizontal 
direction (maroon line).

In all three
cases we observe that the DM signal for a hit in vertical direction
has again three peaks
corresponding to three eigenfrequencies, which occur also in the vertical
thermal noise. Interestingly not in all cases we see a clear resonance shape,
but we also see the anti-resonance-like behaviour as discussed in
Sec.~\ref{sec:Cancellation}, for instance, for the second peak in the
TM case, in Fig.~\ref{fig:TMKagra}.
For hits in horizontal direction, we also observe that all resonance
frequencies get excited and especially for hits in the BS and IM
we see many cancellations although most of them are not around 
the peaks.
These differences in the spectrum could be used 
as a way to distinguish DM hits from background noise as we 
mentioned before. Another way is to look for a daily or annual 
modulation effect \cite{Drukier:1986tm, Collar:1992qc} due to 
the motion of the Earth with respect to the DM halo. That leads 
to a modulation of the hit rate but also preferred recoil 
direction which could be resolved with enough statistics.

In general, we see that the DM signal is much smaller than the noise.
For instance, for the hit in the TM the ratio of the DM induced strain 
in vertical (horizontal) direction 
over the noise strain is always smaller than $10^{-10}$ ($10^{-7}$) 
over the considered frequency range. 
With such small figures, the detection of light particle DM 
in KAGRA seems rather impossible. For LIGO and other 
earth-bound GW detectors we expect similar
results since the mirrors all have similar masses and hence the recoil 
energy is usually much smaller than the thermal energy.
Very heavy DM or DM boosted in some way to relativistic velocities can 
help to overcome this suppression but the number densities for both 
cases is usually expected to be small making any scattering 
event seem unlikely.

\subsection{SNR}
\label{sec:KAGRASNR}

Although just looking at the DM induced strain compared
to the noise strain a detection seems rather hopeless,
we discuss the SNR still in some detail to comment on some 
insights.
We will use again basically the same formula as before
\begin{equation}
\varrho^2 = \int_{f_\text{min}}^{f_\text{max}} \diff f \frac{4 \, |\tilde{x}_{\text{DM}}(2 \, \pi \, f)|^{2}}{S_{\text{tot}}(2 \, \pi \, f)} \;.
\end{equation}
A major difference is that the noise in this case is much more
complicated and also the DM signal shows more than one peak.
Still we want to focus on two peaks, the second peak of the 
vertical noise at $f_v = 31.4$~Hz and the first peak of the 
horizontal noise at $f_h = 175.4$~Hz. The quantum noise for these 
two peaks is comparatively small, cf.~Fig.~\ref{fig:KAGRAnoises}. 
For the integration boundaries
we could in theory again use the FWHM of the peaks, which is nevertheless
probably smaller than the KAGRA frequency resolution. Therefore, we use 
more conservative boundaries
\begin{equation} 
 f_{\text{min},v} = 30.4 ~\text{Hz} \text{ and }  f_{\text{max},v} = 32.4 ~\text{Hz}
\end{equation} 
for the vertical suspension noise peak and
\begin{equation} 
 f_{\text{min},h} = 170 ~\text{Hz} \text{ and }  f_{\text{max},h} = 180 ~\text{Hz}
 \label{eq:horizontalboundary}
\end{equation} 
for the horizontal suspension noise peak. Within these boundaries
the thermal suspension noise is still dominant over the other noise
components. We then get two SNR values, $\varrho^2_v$ and
$\varrho^2_h$, where the index refers to the integration boundaries, i.e.
\begin{equation}
 \varrho_h^2 = \int_{f_\text{min},h}^{f_\text{max},h} \diff f \frac{4 \, |\tilde{x}_{\text{DM}}(2 \, \pi \, f)|^{2}}{S_{\text{tot}}(2 \, \pi \, f)} \;.
\end{equation}

For a DM hit in the TM in vertical direction
(where  $|\tilde{x}_{\text{DM}}|^2 = \text{VHC}^2 |\tilde{x}_{\text{DM},v}|^2 $)
using our standard assumptions
we then find $\varrho^2_v = 1.89 \times 10^{-21}$
and $\varrho^2_h = 1.73 \times 10^{-22}$
While for a DM hit in horizontal direction
(where  $|\tilde{x}_{\text{DM}}|^2 = |\tilde{x}_{\text{DM},h}|^2 $)
we find 
$\varrho^2_v = 1.47 \times 
10^{-17}$ and $\varrho^2_h = 6.94 \times 10^{-18}$.
Taking the ratio $\varrho^2_v/\varrho^2_h$ we can extract
some information on the direction of the recoil momentum.
In  the extreme cases we find $\varrho^2_v/\varrho^2_h = 10.9$
for a hit in purely vertical direction while for a hit in purely
horizontal direction we find $\varrho^2_v/\varrho^2_h = 2.1$.
These numbers nevertheless suffer from ambiguities. We cannot
reconstruct the sign of the recoil momentum, cf.~eq.~\eqref{eq:vTMDM},
and we also cannot reconstruct in which of the two KAGRA arms
the DM hit occurs. In a dedicated DM experiment based on an
interferometer it might be useful to slightly detune the different 
components of the experiment to resolve these ambiguities.

The value for the SNR we computed here for a DM hit in 
horizontal direction, $\varrho^2_h = 6.94 \times 10^{-18}$,
is different from the one in the toy model for the same peak, 
see Sec.~\ref{subsec:toymodelSNR} 
where $\varrho^2_{\text{th}} = 4.09 \times 10^{-24}$. There 
are three main reasons for that. First, we do not use the 
same boundaries here and there. Even if we would apply the 
same boundaries from eq.~\eqref{eq:horizontalboundary} to the toy 
model for computing the SNR see eq.~\eqref{eq:SNRth}, we would find 
$\varrho^2_{\text{toy}} = 3.68 \times 10^{-14}$ which is still 
different from the value we have here.
That is due to the fact, that here  
$|\tilde{x}_{\text{DM}}|^2/( f \, S_{\text{tot}} ) $
is not a constant as in the toy model. We can understand this 
since here
\begin{align}
 S_{\text{tot}} \approx S_{\text{th},h/v} \sim \Im(D_{v/h}^{-1})_{33} \;, 
\end{align}
while
\begin{align}
 |\tilde{x}_{\text{DM}}|^2 \sim | (D_{v/h}^{-1})_{33} |^2 \;.
\end{align}
In the case of the one-dimensional single oscillator in the toy model
these two had the same structure while here this is more complicated.
In fact, the thermal suspension noises $S_{\text{th},h/v}$ do not have
the shape of a single Lorentzian function anymore. Also note that $|\tilde{x}_{\text{DM}}|^2$
clearly exhibits cancellations which are not described in terms of a sum of
Lorentzian functions emphasizing that this approximation breaks down.
The third reason is that in the toy model section we did not include 
the quantum noise when calculating $\varrho^2_{\text{th}}$ which 
nevertheless would not alter the value as drastically as the 
previous two reasons since it is subdominant within the 
integration boundaries.

Unsurprisingly reality is more complicated
than a toy model. The suppression factor $E_R/E_{\text{th}}$ is nevertheless
still there since it is related to constant factors of $|\tilde{x}_{\text{DM}}|^2$
and $S_{\text{th},h/v}$. To enhance the SNR it therefore still seems most efficient
to us to focus first on enlarging $E_R/E_{\text{th}}$ using lighter and colder mirrors.

\section{Comments on other technologies}
\label{sec:Others}

In this section, we just briefly want to comment on space-based
GW detectors and atomic interferometers, which
are very different setups compared to conventional
terrestrial laser interferometers and hence our previously derived
results cannot be immediately translated.

\subsection{Space-based GW Detectors}
\label{sec:Space} 

Let us begin with space-based GW detectors
where the TMs should be ideally free-falling.
Our approximation as damped harmonic oscillators is then 
inappropriate. In fact, in LISA pathfinder (LPF) \cite{McNamara:2008zz}
and LISA \cite{Audley:2017drz} the position of the TMs are
unstable due to various sources of noise which needs to be corrected
in a controlled manner.

We can nevertheless, give some rough estimate to see, if light DM
could be detected in these experiments. We follow the approach
in \cite{Thorpe:2015cxa} where they looked for
micrometeroidal events at LPF which hit the spacecraft. In this reference, the relevant
Fourier transform of a force acting on the target mass
for a duration much shorter than the time resolution at $t = 0$
is just $\tilde{F} = q_R$. Then it is also important, that
the target sensitivity in terms of the amplitude spectral density
is given in terms of the differential acceleration of two TMs
as 
\begin{align}
\sqrt{S_{\Delta g} } \leq 3 \sqrt{2} \text{ fm s}^{-2}/\sqrt{\text{Hz}} \times \sqrt{1 + (f/8 \text{ mHz})^4}
\end{align}
within the frequency band of $0.1$~mHz $\leq f \leq 1$~Hz
for LISA and
\begin{align}
\sqrt{S_{\Delta g} } \leq 30 \text{ fm s}^{-2}/\sqrt{\text{Hz}} \times \sqrt{1 + (f/3 \text{ mHz})^4}
\end{align} 
within the frequency band of $1$~mHz $\leq f \leq 30$~Hz for 
LPF, cf.~\cite{Armano:2016bkm}.
LPF in the end exceeded not only the LPF target sensitivity,
but remarkably even the LISA target sensitiviy~\cite{Armano:2018kix}.
That gives the hope that also LISA will do much better than its target sensitivity.

These target sensitivities have to be compared to the
amplitude spectral density induced by a DM hit
\begin{equation}
\sqrt{S_{\Delta g, \text{DM}}} = 2 \sqrt{f} \left| \frac{\tilde{F}}{m_T} \right| = 2 \sqrt{f} \left| \frac{q_R}{m_T} \right| \;.
\end{equation}
This is the analogue to eq.~\eqref{eq:toyDMstrain} with $|\tilde{x}_{\text{DM}}|$
replaced by $|\tilde{F}/m_T|$. Using again as a rough estimate for the recoil momentum
$|q_R| \sim m_{\text{DM}} |\vec{v}_{\text{DM}}|$ with $m_{\text{DM}} = 1$~GeV/c$^2$,
$|\vec{v}_{\text{DM}}| = 220$~km/s and $m_T = 1.928$~kg for LPF we find
\begin{equation}
\sqrt{S_{\Delta g, \text{DM}}} \sim  4.1 \times 10^{-7} \, \sqrt{\frac{f}{\text{Hz}}} \; \text{ fm s}^{-2}/\sqrt{\text{Hz}}  \;.
\end{equation}
From that estimate it is clear that the up-coming space-based GW
detectors are not suited to detect conventional light DM.

We can also use the LPF results \cite{Thorpe:2015cxa} to estimate
the expected SNR
\begin{equation}
\varrho^{2}_{\text{LPF}} = \frac{|q_R|}{P_c} \sim \frac{m_{\text{DM}} |\vec{v}_{\text{DM}}|}{P_c} \approx 10^{-14} \;,
\end{equation}
where $P_c \approx 3.6 \times 10^{-8}$~N~s~\cite{Thorpe:2015cxa}. Based on this
simple estimation, one can compare the SNR obtained here with the one from the
toy model (with the typical KAGRA parameters) discussed in subsection 2.3.
Interestingly, LPF is much better than our toy KAGRA setup
$\varrho^{2}_{\text{LPF}} \approx 10^{-14} \gg
\varrho^{2}_{\text{toy}} \approx \varrho^{2}_{\text{th}} \approx 10^{-24}$.
That suggests that a light DM detector
based on interferometers might be more easily realised in space than on earth.

\subsection{Atomic Interferometers}
\label{sec:Atoms} 

Recently, there have also been some proposals to use atomic interferometers
for GW detection instead of laser interferometers, for instance, AGIS~\cite{Dimopoulos:2008sv}, 
MAGIS~\cite{Graham:2017pmn,Coleman:2018ozp},
AION~\cite{Badurina:2019hst},
AEDGE~\cite{Bertoldi:2019tck}, MIGA~\cite{Canuel:2017rrp},
ELGAR~\cite{Canuel:2019abg}, ZAIGA~\cite{Zhan:2019quq}
see also \cite{Dimopoulos:2008hx,Graham:2012sy,Graham:2016plp}.
Many of these proposals also mention the possibility to detect wave-like, ultralight
DM with masses well below 1~eV/c$^2$. The signal would then be wave-like as
well with a frequency determined by DM parameters quite different from the
case of scattering of individual DM particles as we discuss here.

These experiments promise remarkable sensitivities. For instance,
MAGIS-100, might be able to test signals with an amplitude spectral
density 
down to $10^{-16} \, g/\sqrt{\text{Hz}}$~\cite{Coleman:2018ozp},
cf.~also~\cite{Graham:2015ifn}, which
would be about $0.1 \text{ fm s}^{-2}/\sqrt{\text{Hz}}$
exceeding LISA requirements. Interestingly, the target mass is very
small in these
cases. Atomic interferometers consist of clouds of ultracold atoms. In the
case of MAGIS-100 they mention fluxes of the order of $10^8$~strontium atoms per
second or $10^{15}$ dropped atoms in a year much less than a milligram.
Common strontium isotopes have
an atomic weight of about 87-88~GeV/c$^2$ allowing for much larger
recoil energies than in conventional GW detectors that could be observed
in atomic interferometers.

On the other hand, since the total target mass is so small the number of expected
DM scattering events is very small as well. That is a disadvantage 
compared to more conventional detectors.

\section{Summary and Conclusions}
\label{sec:Summary}

The motivation for this work is to describe scattering of DM
particles at GW detectors.
To understand the physics principles behind it, we studied first a damped harmonic
oscillator with internal damping as simplified toy model. In this model,
it became already clear that DM can excite mechanical resonances allowing for
comparatively large strain values in some frequency windows, which could
help to overcome quantum noise, for instance. The catch is that these
resonances are also excited by thermal energy.
In conventional GW detectors a detection of light DM becomes rather hopeless.
But from our toy model, we can easily understand and quantify
that there are two things that can
be done for improvement and which form
one of the most important results of our paper. First, the temperature could be further reduced since
thermal noise is proportional to it. Active thermal noise
suppression might also help in the future.
Second, the oscillator
mass should be reduced to increase the recoil energy from the 
hit. That two measures would help
to overcome thermal and quantum noise.

We then demonstrate how easy our formalism can be applied to the
much more complicated but realistic example of the KAGRA detector.
The resonance pattern gets much
more involved since there are multiple mechanical resonances that can be excited,
and we also encounter partial cancellations which provide a potential way
to distinguish a DM signal from thermal noise.
Unfortunately,
the expected SNR in KAGRA is very small for light DM on which we focussed here.

We also comment briefly on space-based GW detectors and atomic interferometers.
LISA and LPF turn out to be more sensitive for our physics case, but still not sensitive enough
for a realistic detection. Atomic interferometers are actually promising since the target material
in this case are rather few, ultracold atoms fulfilling our criteria.

In summary, conventional GW detectors have impressive
sensitivities but they are not good particle DM detectors. 
Nevertheless, mechanical sensors at the quantum limit like
KAGRA have recently been more seriously considered as potential
technology for DM detection, cf.~\cite{Carney:2020xol}.
To understand what the ultimate sensitivity
of this new technologies will be, further research is needed
and being carried out at that very moment.
We think that at least in some mass regions such technologies
will become competitive to more conventional approaches
and a good modeling of the DM signal as discussed in this work
will be useful to separate it from irreducible backgrounds
for which we have also presented some ideas.

\section*{Acknowledgment}
\label{sec:Acknowledgment}

CHL and MS are supported by the Ministry of Science and Technology (MOST) of Taiwan 
under grant number MOST 107-2112-M-007-031-MY3.

\end{document}